\documentstyle[prb,aps,floats,epsfig,amssymb,amsmath]{revtex}

\begin{document}
\twocolumn[\hsize\textwidth\columnwidth\hsize\csname@twocolumnfalse
\endcsname
\title{Introduction to the Bethe Ansatz I} 
\author{Michael Karbach$^*$ and Gerhard M\"uller$^\dagger$} 
\address{$^*$Bergische Universit\"at Wuppertal, Fachbereich Physik, 
         D-42097 Wuppertal, Germany \\ 
         $^\dagger$Department of Physics, University of Rhode Island,
         Kingston RI 02881-0817}
\date{\today}
\maketitle
\begin{abstract}
\end{abstract}
\twocolumn
]
A few years after the formulation of quantum mechanics, Heisenberg and
Dirac\cite{Heis26} discovered that one of its wondrous consequences was the key
to the age-old mystery of ferromagnetism.  They found that the laws of quantum
mechanics imply the existence of an effective interaction, $J_{ij}\, {\bf
  S}_i\cdot{\bf S}_j$, between electron spins on neighboring atoms with
overlapping orbital wave functions. The exchange interaction, as it has become
known, is caused by the combined effect of the Coulomb repulsion and the Pauli
exclusion principle. This spin interaction was soon recognized to be the key to
a microscopic theory of ferromagnetism and many other cooperative phenomena
involving electron spins.

In 1931 Hans Bethe\cite{Beth31} presented a method for obtaining the exact
eigenvalues and eigenvectors of the one-dimensional (1D) spin-1/2 Heisenberg
model, a linear array of electrons with uniform exchange interaction between
nearest neighbors.  Bethe's parametrization of the eigenvectors, the {\it Bethe
  ansatz}, has become influential to an extent not imagined at the time. Today,
many other quantum many body systems are known to be solvable by some variant of
the Bethe ansatz, and the method has been generalized and expanded far beyond
the ad hoc calculational tool it was originally. Unlike the simulation of a
classical model system, most computational approaches to quantum many-body
systems require a fair amount of analytical work up front. This requirement is
true for Monte Carlo calculations, renormalization group approaches, the
recursion method, large-scale numerical diagonalizations, and for the Bethe
ansatz when used in a computational context.

In spite of its proven importance and wide range of applications, the Bethe
ansatz is rarely discussed in textbooks on quantum mechanics and statistical
mechanics except at the advanced level.  The goal of this column is to introduce
the Bethe ansatz at an elementary level. In future columns we plan to discuss
some of the extensions, generalizations, and applications of the method, to
which many workers have made important contributions. Our emphasis will be on
computational applications for which the method is less well known than for its
usefulness as an analytical tool.

The Bethe ansatz is an exact method for the calculation of eigenvalues and
eigenvectors of a limited but select class of quantum many-body model systems.
Although the eigenvalues and eigenvectors for a finite system may be obtained
with less effort from a brute force numerical diagonalization, the Bethe ansatz
offers two important advantages: (i) all eigenstates are characterized by a set
of quantum numbers which can be used to distinguish them according to specific
physical properties; (ii) in many cases the eigenvalues and the physical
properties derived from them can be evaluated in the thermodynamic limit.

The Hamiltonian of the Heisenberg model of spins ${\bf S}_n=(S_n^x,S_n^y,S_n^z)$
with quantum number $s=1/2$ on a 1D lattice of $N$ sites with periodic boundary
conditions ${\bf S}_{N+1}={\bf S}_1$ is given by
\begin{eqnarray}\label{H}
  H &=& -J\sum_{n=1}^N {\bf S}_n \cdot {\bf S}_{n+1} \nonumber \\
    &=& -J\sum_{n=1}^N \biggl [
   \frac{1}{2}\bigl( S_n^+S_{n+1}^- + S_n^-S_{n+1}^+ \bigr )
     +S_n^zS_{n+1}^z \biggr ],
\end{eqnarray}
where $S_n^\pm\equiv S_n^x\pm i S_n^y$ are spin flip operators. $H$ acts on a
Hilbert space of dimension $2^N$ spanned by the orthogonal basis vectors
$|\sigma_1 \ldots \sigma_N\rangle$, where $\sigma_n=\uparrow$ represents an up
spin and $\sigma_n=\downarrow$ a down spin at site $n$.  The spin commutation
relations (with $\hbar=1$) are
\begin{equation}\label{SSS}
 [S_n^z,S_{n'}^\pm]=\pm S_n^\pm\delta_{n n'}, \quad
 [S_n^+,S_{n'}^-] = 2 S_n^z \delta_{nn'}.
\end{equation}

The application of the operators $S_n^\pm,S_n^z$ on a vector $|\sigma_1 \ldots
\sigma_N\rangle$ yields the results summarized in Table~\ref{tabvec}. We can use
these rules to express $H$ as a real and symmetric $2^N\times 2^N$ matrix whose
eigenvectors can be computed by standard diagonalization algorithms \cite{GL93}.
From the eigenvectors, the physical quantities of interest can be calculated by
evaluating the expectation values (matrix elements) for the corresponding
operators.

\begin{table}[ht]
\caption{Rules governing the application of the spin operators on
the basis vectors $|\sigma_1\ldots\sigma_N\rangle$ with
$\sigma_n=\uparrow,\downarrow$.}
\begin{tabular}{lcc}\label{tabvec}
  &$|\ldots\uparrow\ldots\rangle$ &
$|\ldots\downarrow\ldots\rangle$
\\
\hline
$S_k^+$& 0                       &
$|\ldots\uparrow\ldots\rangle
$ \\
$S_k^-$&$|\ldots\downarrow\ldots\rangle$& 0 \\
$S_k^z$&\hspace*{-2mm}$\frac{1}{2}|\ldots\uparrow\ldots\rangle$&
\hspace*{-5mm}$-\frac{1}{2}|\ldots\downarrow\ldots\rangle$
\end{tabular}
\end{table}

The Hamiltonian matrix can be written in block diagonal form if we perform,
prior to the numerical diagonalization, one or several basis transformations
from \{$|\sigma_1 \ldots \sigma_N\rangle$\} to a symmetry-adapted basis. These
transformations reduce the computational effort needed for the remaining
numerical diagonalization and make it possible to handle larger system sizes.

The Bethe ansatz is a basis transformation that does not have to be supplemented
by a numerical diagonalization. In principle and usually in practice, this
alternative procedure removes the cap on system sizes. However, the
implementation of the Bethe ansatz entails calculational challenges of its own.
Depending on the specifics of the application, they can be met by analytical or
computational methods as we shall see in the following.

For the Heisenberg model, two symmetries are essential for the application of
the Bethe ansatz. The rotational symmetry about the $z$-axis in spin space,
which we have chosen to be the quantization axis, implies that the $z$-component
of the total spin $S_T^z\equiv\sum_{n=1}^N S_n^z$ is conserved: $[H,S_T^z]=0$.
According to the rules of Table \ref{tabvec}, the operation of $H$ on
$|\sigma_1\ldots\sigma_N\rangle$ yields a linear combination of basis vectors,
each of which has the same number of down spins.  Hence, sorting the basis
vectors according to the quantum number $S_T^z=N/2-r$, where $r$ is the number
of down spins, is all that is required to block diagonalize the Hamiltonian
matrix.

The block with $r=0$ (all spins up) consists of a single vector $|F\rangle
\equiv |\uparrow \ldots \uparrow\rangle$. It is an eigenstate,
$H|F\rangle=E_0|F\rangle$, with energy $E_0=-JN/4$.  The $N$ basis vectors in
the invariant subspace with $r=1$ (one down spin) are labeled by the position of
the flipped spin:
\begin{equation}\label{vecn}
 |n\rangle= S_n^-|F\rangle \qquad n=1,\ldots,N.
\end{equation}
To diagonalize the $r=1$ block of $H$, which has size $N\times N$, we take into
account the translational symmetry, i.e., the invariance of $H$ with respect to
discrete translations by any number of lattice spacings. Translationally
invariant basis vectors can be constructed from the vectors in (\ref{vecn}) by
writing

\begin{equation}\label{sws} 
 |\psi\rangle = \frac{1}{\sqrt{N}}\sum_{n=1}^N e^{ikn} |n\rangle
\end{equation}
for wave numbers $k=2\pi m/N,\; m=0,\ldots,N-1$. (The lattice spacing has been
set equal to unity.)  The vectors $|\psi\rangle$ are eigenvectors of the
translation operator with eigenvalues $e^{ik}$ and are also eigenvectors of $H$
with eigenvalues
\begin{equation}\label{e1k}
 E-E_0=J(1-\cos k),
\end{equation}
as can be verified by inspection (see Problem 1). The vectors (\ref{sws})
represent magnon excitations, in which the complete spin alignment of the
ferromagnetic ground state $|F\rangle$ is periodically disturbed by a spin wave
with wavelength $\lambda=2\pi/k$ (see Problem 2).

In the invariant subspaces with $2\leq r \leq N/2$, the translationally
invariant basis does not completely diagonalize the Hamiltonian matrix even if
we take into account further symmetries of (\ref{H}) such as the full rotational
symmetry in spin space or the reflection symmetry on the lattice. Here the Bethe
ansatz is a powerful alternative.

For $r = 1$, the case we have already solved, we now proceed somewhat
differently. Any eigenvector in the $r=1$ subspace is a superposition of the
basis vectors (\ref{vecn}):
\begin{equation}\label{psin}
 |\psi\rangle = \sum_{n=1}^N a(n)|n\rangle.
\end{equation}
The eigenvector $|\psi\rangle$ is a solution of the eigenvalue equation
$H|\psi\rangle=E|\psi\rangle$ if the coefficients $a(n)$ satisfy the linear
equations
\begin{equation}\label{e1}
 2[E-E_0]a(n)=J[2a(n)-a(n-1)-a(n+1)]
\end{equation}
for $n=1,2,\ldots,N$ and with periodic boundary conditions $a(n+N) = a(n)$.  $N$
linearly independent solutions of (\ref{e1}) are
\begin{equation}\label{an}
 a(n)=e^{ikn}, \qquad k=\frac{2\pi}{N}m, \qquad m=0,1,\ldots,N-1.
\end{equation}
If we substitute the coefficients $a(n)$ into (\ref{psin}), we obtain (after
normalization) the magnon states (\ref{sws}) with energies (\ref{e1k}).

The distinctive features of the Bethe ansatz begin to emerge when we apply the
same procedure to the case $r=2$. The task is to determine the coefficients
$a(n_1,n_2)$ for all eigenstates
\begin{equation}\label{psi2}
 |\psi\rangle = \! \sum_{1\leq n_1<n_2\leq N} a(n_1,n_2)
|n_1,n_2\rangle,
\end{equation}
where $|n_1,n_2\rangle\equiv S_{n_1}^-S_{n_2}^-|F\rangle$ are the basis vectors
in this subspace of dimension $N(N-1)/2$. Bethe's preliminary ansatz for the
coefficients is
\begin{equation}\label{a1a2}
 a(n_1,n_2) = A e^{i(k_1n_1+k_2n_2)}+A'e^{i(k_1n_2+k_2n_1)}.
\end{equation}
We might wish to set $A=A'$, use the same values for $k_1,k_2$ as in (\ref{an}),
and interpret the wave function as a superposition of two magnons. However, the
result would be an overcomplete set of $N(N+1)/2$ nonorthogonal and
nonstationary states. Superimposed spin waves are in conflict with the
requirement that the two flipped spins must be at different sites. The
eigenvalue equation for (\ref{psi2}) translates into $N(N-1)/2$ equations for as
many coefficients $a(n_1,n_2)$:
\begin{mathletters}\label{con2}
\begin{eqnarray}
 2[E-E_0] a(n_1,n_2) &=& J[4a(n_1,n_2)\!-\!a(n_1\!-\!1,n_2) 
 \nonumber \\  && \hspace*{-2.5cm} 
 -a(n_1\!+\!1,n_2) -a(n_1,n_2\!-\!1)-a(n_1,n_2\!+\!1)] 
\nonumber \\ && \hspace*{-2.5cm} \mbox{for} \quad n_2>n_1\!+\!1 \label{b1},
\\
 2[E-E_0] a(n_1,n_2) &=& J[2a(n_1,n_2)-a(n_1\!-\!1,n_2)
 \nonumber \\  && \hspace*{-2.5cm} -a(n_1,n_2\!+\!1)]
\nonumber \\ && \hspace*{-2.5cm} \mbox{for} \quad n_2=n_1\!+\!1. \label{b2}
\end{eqnarray}
\end{mathletters}
Equations (\ref{b1}) are satisfied by $a(n_1,n_2)$ in (\ref{a1a2}) with
arbitrary $A,A',k_1,k_2$ for $n_2>n_1+1$ and for $n_2=n_1+1$ provided the energy
depends on $k_1,k_2$ as follows:
\begin{equation}\label{Ek}
 E -E_0 = J\! \sum_{j=1,2}(1-\cos k_j).
\end{equation}
Equations~(\ref{b2}), which are not automatically satisfied by the ansatz
(\ref{a1a2}), are then equivalent to the $N$ conditions
\begin{equation}\label{ann}
 2a(n_1,n_1+1)=a(n_1,n_1)+a(n_1+1,n_1+1)
\end{equation}
obtained by subtracting (\ref{b2}) from (\ref{b1}) for $n_2=n_1+1$. In other
words, $a(n_1,n_2)$ are solutions of Eqs.~(\ref{con2}) if they have the form
(\ref{a1a2}) and satisfy (\ref{ann}). The conditions (\ref{ann}) require a
modification of the amplitude ratio,
\begin{equation}\label{AA}
 \frac{A}{A'}\equiv e^{i\theta}=
  -\frac{e^{i(k_1+k_2)}+1-2e^{ik_1}}{e^{i(k_1+k_2)}+1-2e^{ik_2}}.
\end{equation}
This requirement is incorporated into the Bethe ansatz as extra phase factors
\begin{eqnarray}\label{ba}
 a(n_1,n_2) = e^{i(k_1n_1+k_2n_2+\frac{1}{2}\theta_{12})}
  + e^{i(k_1n_2+k_2n_1+\frac{1}{2}\theta_{21})},
\end{eqnarray}
where the phase angle $\theta_{12}=-\theta_{21}\equiv \theta$ depends on the as
yet undetermined $k_1,k_2$ via (\ref{AA}) or, in real form, via
\begin{equation}\label{ba2}
 2\cot\frac{\theta}{2} = \cot\frac{k_1}{2} - \cot\frac{k_2}{2}.
\end{equation}
The quantities $k_1,k_2$ will henceforth be referred to as momenta of the Bethe
ansatz wave function (\ref{psi2}) with coefficients (\ref{ba}).

Two additional relations between $k_1,k_2,$ and $\theta$ follow from the
requirement that the wave function (\ref{psi2}) be translationally invariant,
which implies that $a(n_1,n_2)=a(n_2,n_1+N)$. This condition is satisfied by the
coefficients (\ref{ba}) if $e^{ik_1N} = e^{i\theta},\, e^{ik_2N} =
e^{-i\theta}$.  Equivalently, we can write (after taking logarithms)
\begin{equation}\label{bapbc2}
 Nk_1 = 2\pi\lambda_1 + \theta, \qquad
 Nk_2 = 2\pi\lambda_2 - \theta,
\end{equation}
where the integers $\lambda_i\in\{0,1,\ldots,N-1\}$ are called Bethe quantum
numbers.

The remaining task is to find all $(\lambda_1,\lambda_2)$ pairs which yield
solutions of Eqs.~(\ref{ba2}) and (\ref{bapbc2}), known as the Bethe ansatz
equations. Every eigenstate in the $r=2$ subspace can be found in this way. For
any solution $k_1,k_2,\theta$, the (non-normalized) eigenvector has coefficients
of the form (\ref{ba}). The expression (\ref{Ek}) for the energy and the
relation
\begin{equation}\label{k12}
 k = k_1+k_2= \frac{2\pi}{N}(\lambda_1+\lambda_2)
\end{equation}
for the wave number $k$ are reminiscent of two superimposed magnons.  The magnon
interaction is reflected in the phase shift $\theta$ and in the deviation of the
momenta $k_1,k_2$ from the values of the one-magnon wave numbers as given in
(\ref{an}).  We shall see that the magnons either scatter off each other or form
bound states. Note that the momenta $k_1,k_2$ specify the Bethe ansatz wave
function (\ref{psi2}), while the wave number $k$ is the quantum number
associated with the translational symmetry of $H$ and exists independently of
the Bethe ansatz.

The analysis of the complete $r=2$ spectrum will demonstrate the usefulness of
the Bethe quantum numbers for distinguishing eigenstates with different physical
properties. The allowed $(\lambda_1,\lambda_2)$ pairs are restricted to
$0\leq\lambda_1\le\lambda_2\leq N-1$. Switching $\lambda_1$ and $\lambda_2$
simply interchanges $k_1$ and $k_2$ and produces the same solution.  There are
$N(N+1)/2$ pairs that meet this restriction, but only $N(N-1)/2$ of them yield a
solution of Eqs.~(\ref{ba2}) and (\ref{bapbc2}). The solutions can be determined
analytically or computationally. Some of them have real $k_1,k_2$, and others
yield complex conjugate momenta, $k_2=k_1^*$.

We first find all the solutions and then interpret them. The
$(\lambda_1,\lambda_2)$ pairs which yield solutions for $N=32$ are shown in
Fig.~1.
\begin{figure}[htb]
  \centerline{\epsfig{file=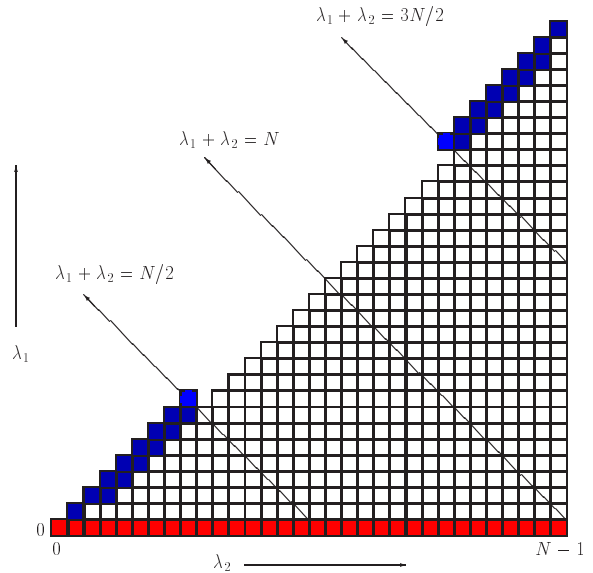,width=9cm}}
  \vspace{-3cm}
  \caption[1]{The allowed pairs of Bethe quantum numbers $(\lambda_1,\lambda_2)$
    that characterize the $N(N-1)/2$ eigenstates in the $r=2$ subspace for
    $N=32$.  The states of class $C_1,C_2$, and $C_3$ are colored red, white,
    and blue, respectively.}
  \label{FIG_Bethenumbers}
\end{figure}
We begin with the class $C_1$ of states for which one of the Bethe quantum
numbers is zero, $\lambda_1=0,\lambda_2=0,1,\ldots,N-1$. There exists a real
solution for all $N$ combinations, $k_1=0,k_2=2\pi\lambda_2/N,\theta=0$ (see
Problem 3).

Next consider the class $C_2$ of states with nonzero $\lambda_1,\lambda_2$ which
differ by two or more: $\lambda_2-\lambda_1\geq 2$. There are $N(N-5)/2+3$ such
pairs. All of them yield a solution with real $k_1,k_2$. To determine these
solutions, we combine Eqs.~(\ref{ba2}), (\ref{bapbc2}), and (\ref{k12}) into a
single nonlinear equation for $k_1$:
\begin{equation}\label{r2ss}
 2\cot \frac{N k_1}{2} = \cot \frac{k_1}{2} - \cot \frac{k-k_1}{2}.
\end{equation}
The solution $k_1$ of (\ref{r2ss}) for a given wave number $k=(2\pi n/N),\;
n=0,1,\ldots,N-1$ is substituted into (\ref{bapbc2}) and yields $k_2$ and
$\theta$. Some of the solutions can be found analytically (Problem 4), and
others must be determined numerically (Problem 5).

The remaining class $C_3$ of states has nonzero Bethe quantum numbers
$\lambda_1, \lambda_2$ which either are equal or differ by unity.  There exist
$2N-3$ such pairs, but we will see that only $N-3$ pairs yield solutions of
(\ref{ba2}) and (\ref{bapbc2}). Most of the class $C_3$ solutions are complex.
To find them, we write
\begin{equation}\label{k1k2c}
 k_1\equiv \frac{k}{2}+iv, \quad
 k_2\equiv \frac{k}{2}-iv, \quad
 \theta\equiv\phi+i\chi,
\end{equation}
and use Eqs.~(\ref{ba2}) and (\ref{bapbc2}) for fixed $k$ to obtain the relation
\begin{equation}\label{r2bs}
 \cos\frac{k}{2} \sinh (Nv) = \sinh [(N-1)v] + \cos\phi \sinh v,
\end{equation}
where $\phi=\pi(\lambda_1-\lambda_2)$, and $\chi=N v$ is inferred from the
solution.  It is sufficient to consider $v>0$. The energy (\ref{Ek}) of any
complex solution is rewritten in the form
\begin{equation}\label{Ekc}
 E-E_0 = 2J \biggl(1-\cos \frac{k}{2} \cosh v \biggr).
\end{equation}
Inspection of (\ref{r2bs}) shows (see Problem 6) that a complex solution exists
for $\lambda_2=\lambda_1$ if
\begin{equation}\label{l1eql2}
 \lambda_1\!+\!\lambda_2=2,4,\ldots,\frac{N}{2}\!-\!2,\frac{3N}{2}\!+\!2,
\ldots,2N\!-\!2 .
\end{equation}
For $\lambda_2=\lambda_1+1$ a complex solution exists if
\begin{equation}
 \lambda_1\!+\!\lambda_2=
\tilde{\lambda},\tilde{\lambda}\!+\!2,\ldots,N/2\!-\!1,3N/2\!+\!3,\ldots,2N
-\tilde{\lambda}\!+\!2,
\end{equation}
where $\tilde{\lambda}\approx\sqrt{N}/\pi$.  In the latter case, there exist
additional real solutions of (\ref{r2ss}) if
\begin{equation}\label{l1l2com2}
   \lambda_1+\lambda_2=3,5,\ldots,\tilde{\lambda}-2,2N-\tilde{\lambda}
   +2,\ldots,2N-3.
\end{equation}
The ($\lambda_1,\lambda_2$) pairs in (\ref{l1eql2})--(\ref{l1l2com2}) account
for $N-4$ solutions in class $C_3$. There is one more class $C_3$ solution. This
state, which has $k=\pi$ and $\lambda_1=\lambda_2=N/4$, is easily missed in the
numerical analysis, because $k_1,k_2$ have an infinite imaginary part (see
Problem 7).

The complete $r=2$ excitation spectrum $(E-E_0)/J$ versus $k$ for a system with
$N=32$ spins as obtained by the analytical and computational procedures outlined
in Problems 3--7 is plotted in Fig.~2.  The $N$ states of class $C_1$ form a
branch with exactly the same dispersion relation (\ref{e1k}) as the magnon
states that populate the $r=1$ subspace. This degeneracy is a consequence of the
conservation of the total spin $S_T$ (see Problem 8). The class $C_2$ states are
spread in a regular pattern over a region in $(k,E)$-space. They are nearly free
superpositions of two one-magnon states.

\begin{figure}[ht]\label{FIG_Eversusk}
  \centerline{\epsfig{file=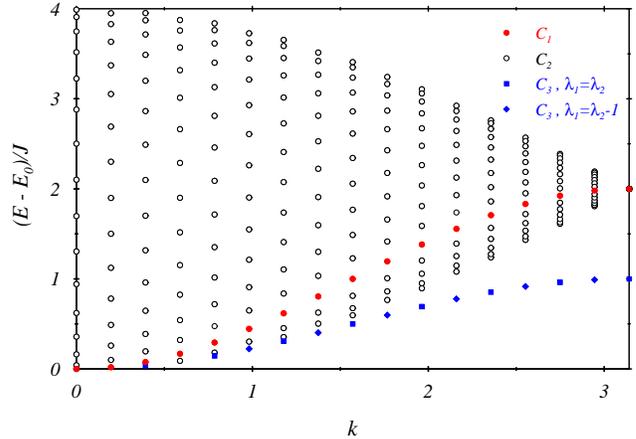,width=7cm,angle=-90}}
  \caption[]{Excitation energy $(E-E_0)/J$ versus wave number $k$
  of all $N(N-1)/2$ eigenstates in the invariant subspace with $r=2$ overturned
  spins for a system with $N=32$. States of class $C_1$ are denoted by red
  circles, states of class $C_2$ by open black circles, and states of class
  $C_3$ by blue squares if $\lambda_2=\lambda_1$, or blue diamonds if
  $\lambda_2=\lambda_1+1$.}
\end{figure}

The excitations belonging to classes $C_1$ and $C_2$ can be characterized as
two-magnon scattering states. The effect of the magnon interaction on these
states is visualized in Fig.~3. It shows all $N(N-1)/2-N+3$ class $C_1$ and
class $C_2$ states for $N=32$ in comparison with the $N(N+1)/2$ two-magnon
superpositions, where the momenta $k_j,j=1,2$ in (\ref{Ek}) are replaced by
one-magnon wave numbers $k_j=2\pi m_j/N,\; m_j=0,1,\ldots,N-1$.  The magnon
interaction manifests itself as a modified excitation energy of the two-magnon
scattering states. Note that the interaction energy approaches zero when either
$k_1$ or $k_2$ goes to zero. The class $C_1$ states can then be interpreted as
exact superpositions of a $k_1=0$ magnon and a $k_2\neq 0$ magnon.

\begin{figure}[ht]\label{FIG_Eversuskall}
  \centerline{\epsfig{file=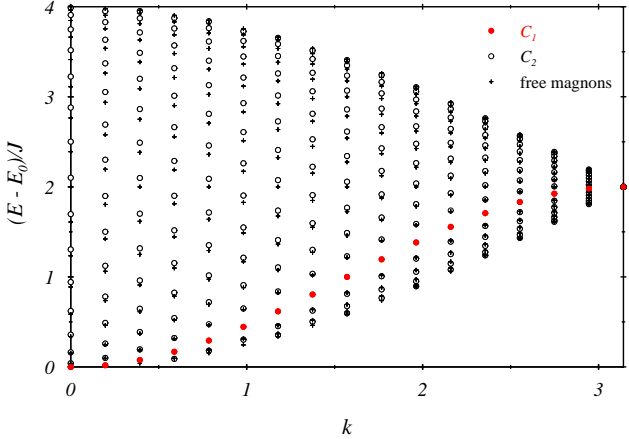,width=7cm,angle=-90}}
  \caption[]{Excitation energy $(E-E_0)/J$ versus wave number
  $k$ of all two-magnon scattering states (classes $C_1$ and $C_2$ from Fig.~2)
  for a system with $N=32$ in comparison with the noninteracting magnon pairs
  (+).}
\end{figure}

As $N$ increases, the energy correction due to the magnon interaction diminishes
for the class $C_2$ states as well and vanishes in the limit $N\to\infty$.  The
two-magnon scattering states then form a continuum with boundaries $E-E_0 = 2J
(1\pm \cos k/2)$, which coincide with those of the continuum of free two-magnon
states.

For the class $C_3$ states, the effects of the magnon interaction are much more
prominent, and they do not disappear in the limit $N\to\infty$. In the
$(k,E)$-plane of Fig.~2, these states lie on a single branch with dispersion
(see Problem 9)
\begin{equation}\label{EkcNinfty}
 E-E_0 = \frac{J}{2}(1-\cos k)
\end{equation}
below the continuum of two-magnon scattering states. They are the two-magnon
bound states.

The bound state character of the class $C_3$ states manifests itself in the
enhanced probability that the two flipped spins are on neighboring sites of the
lattice. This property of the wave function is best captured in the weight
distribution $|a(n_1,n_2)|$ of basis vectors with flipped spins at sites $n_1$
and $n_2$. In Fig.~4 we plot $|a(n_1,n_2)|$ versus $n_2-n_1$ for a sequence of
class $C_3$ states between $k=0$ and $k=\pi$. The distribution is peaked at
$n_2-n_1=1$. Its width is controlled by the imaginary parts of $k_1,k_2,\theta$
in the coefficients (\ref{ba}).  The smallest width is observed in the bound
state at $k=\pi$, that is, for $\lambda_1=\lambda_2=N/4$, whose $k_1,k_2$ have
an infinite imaginary part (see Problem 7).  In this case, all coefficients in
(\ref{psi2}) with $n_2\neq n_1+1$ are zero, which implies that the two down
spins are tightly bound together and have the largest binding energy.  For the
adjacent bound state with quantum numbers $\lambda_1=\lambda_2=N/4-1$, the
exponential dependence of the weight $|a(n_1,n_2)|$ on the separation $n_2-n_1$
can be worked out analytically (see Problem 9).

\begin{figure}[ht]\label{FIG_anm}
  \centerline{\hspace*{2cm}\epsfig{file=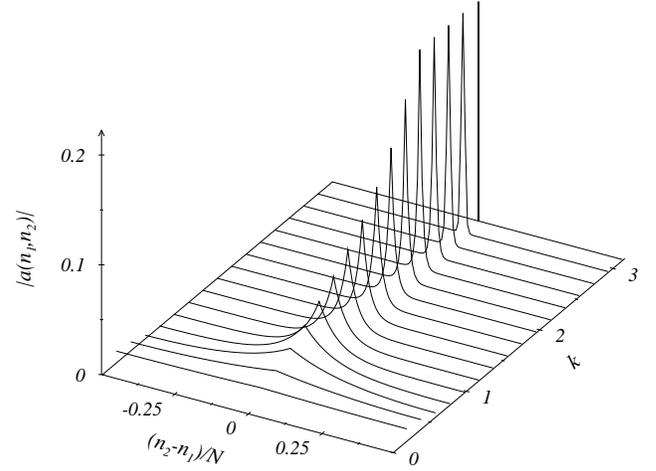,width=8cm,angle=-90}}
  \caption[]{Weight distribution $|a(n_1,n_2)|$ versus distance
  $n_2-n_1$ of the two down spins of class $C_3$ states at
  $k=(2\pi/N)n,\;n=4,8,\ldots,N/2$ for $N=128$.}
\end{figure} 

The width of the distribution $|a(n_1,n_2)|$ increases as $k$ decreases, and the
binding of the two down spins loosens. For finite $N$, the exponential factors
disappear when the distribution has acquired a certain width, and the Bethe
ansatz solutions switch from complex to real. In contrast, the distribution
$|a(n_1,n_2)|$ for scattering states is always broad and oscillates wildly in
general. However, for some combinations of $k_1,k_2$, a smooth distribution
ensues, which has its maximum when the two down spins are farthest apart
($n_2=n_1+N/2$) (see Problem 10).

The formation of bound states and scattering states of specific elementary
excitations exist in a large variety of physical contexts. But only in rare
cases such as this one can the nature and the properties of such compound
excitations be investigated (analytically and computationally) on the level of
detail made possible by the Bethe ansatz.

We now present the Bethe ansatz for an unrestricted number $r$ of overturned
spins. We generalize (\ref{psi2}) and expand the eigenstates in the form
\begin{equation}\label{psir}
 |\psi\rangle = \sum_{1\leq n_1<\ldots<n_r\leq N} a(n_1,\ldots,n_r)
 |n_1,\ldots,n_r\rangle.
\end{equation}
The subspace has dimension $N!/[(N-r)!r!]$. The generalization of (\ref{ba}) for
the coefficients in terms of $r$ momenta $k_j$, and one phase angle
$\theta_{ij}=-\theta_{ji}$ for each $(k_1,k_2)$ pair is as follows:
\begin{eqnarray}\label{bar}
 a(n_1,\ldots,n_r) =
 \sum_{{\cal P}\in S_r}
 \exp && \left( i\sum_{j=1}^r k_{{\cal P} j}n_j
 + \frac{i}{2}\sum_{i<j} \theta_{{\cal P}i{\cal P}j} \right).
\nonumber \\
\end{eqnarray}
The sum ${\cal P}\in S_r$ is over all $r!$ permutations of the labels
$\{1,2,\ldots,r\}$. For $r=2$ the two permutations are the identity $(1,2)$ and
the transposition $(2,1)$, which produce the two terms of (\ref{ba}). The
consistency equations for the coefficients $a(n_1,\ldots,n_r)$ inferred from the
eigenvalue equation $H|\psi\rangle=E|\psi\rangle$ are
\begin{mathletters}\label{conr}
\begin{eqnarray}
   2[E-E_0]a(n_1,\ldots,n_r) &=&
 J\sum_{i=1}^r\sum_{\sigma=\pm1} [a(n_1,\ldots,n_r)
  \nonumber \\ && \hspace*{-2cm}
  \!-\!a(n_1,\ldots,n_i+\sigma,\ldots,n_r)], \label{b1r}
\end{eqnarray}
for $n_{j+1}>n_j+1,\; j=1,\ldots,r$, and
\begin{eqnarray}
  2[E-E_0] a(n_1,\ldots,n_r) &=& \nonumber \\ && \hspace*{-4cm}
  J\hspace*{-4mm}\sum_{i\neq j_\alpha,j_\alpha+1}^r \sum_{\sigma=\pm1}
        [a(n_1,\ldots,n_r) \!-\!a(n_1,\ldots,n_i\!+\!\sigma,\ldots,n_r)]
\nonumber \\ && \hspace*{-4cm} +J\sum_{\alpha}
[2a(n_1,\ldots,n_r)\!-\!a(n_1,\ldots,n_{j_\alpha}
 \!-\!1,n_{j_\alpha\!+\!1},\ldots,n_r)
\nonumber\\ && \hspace*{-2.5cm}
         -a(n_1,\ldots,n_{j_\alpha},n_{j_\alpha+1}+1,\ldots,n_r)] \label{b2r}
\end{eqnarray}
\end{mathletters}
for $n_{j_\alpha+1}=n_{j_\alpha}+1,\; n_{j+1}> n_j+1,\; j\neq j_\alpha$.  The
coefficients $a(n_1,\ldots,n_r)$ are solutions of (\ref{conr}) for the energy
\begin{equation}\label{Ekr}
 E-E_0 = J\sum_{j=1}^r(1-\cos k_j)
\end{equation}
if they have the form (\ref{bar}) and satisfy the conditions
\begin{eqnarray}\label{annr}
 2a(n_1,\ldots\!,n_{j_\alpha}, n_{j_\alpha} \!+\! 1,\ldots\!,n_r) &=&
 \nonumber \\ && \hspace*{-5cm}
 a(n_1,\ldots\!,n_{j_\alpha},n_{j_\alpha},\ldots\!,n_r)
 \!+\!a(n_1,\ldots\!,n_{j_\alpha} \!+\! 1,n_{j_\alpha} \!+\! 1,\ldots\!,n_r)
\nonumber \\
\end{eqnarray}
for $\alpha=1,\ldots,r$ just as we have shown for $r=2$. These conditions relate
every phase angle $\theta_{ij}$ to the (as yet undetermined) $k_j$ in
(\ref{psir}):
\begin{equation}\label{ethetaij}
 e^{i\theta_{ij}}=
 -\frac{e^{i(k_i+k_j)}+1-2e^{ik_i}}{e^{i(k_i+k_j)}+1-2e^{ik_j}} .
\end{equation}
An equivalent relation in real form reads
\begin{equation}\label{ba2r}
 2\cot\frac{\theta_{ij}}{2} = \cot\frac{k_i}{2} - \cot\frac{k_j}{2},
 \quad i,j=1,\ldots,r.
\end{equation}

The translational invariance of (\ref{psir}) implies that the coefficients
(\ref{bar}) satisfy the relation $a(n_1,\ldots,n_r)=a(n_2,\ldots,n_r,n_1+N)$.
Consequently, we must have
\begin{eqnarray}
  \sum_{j=1}^r k_{{\cal P}j}n_j +
  \frac{1}{2} \sum_{i<j}\theta_{{\cal P}i,{\cal P}j}
  &=&
  \frac{1}{2} \sum_{i<j}\theta_{{\cal P}'i,{\cal P}'j} - 2\pi\lambda_{{\cal P}'r}
  \nonumber \\ && \hspace*{-2cm}
  +\sum_{j=2}^r k_{{\cal P}'(j-1)}n_j + k_{{\cal P}'r}(n_1+N),
\end{eqnarray}
where the relation between the permutations on the left and the right is ${\cal
  P}'(i-1)={\cal P}i, \; i=2,\ldots,r;\;{\cal P}'r={\cal P}1$. If we take into
account that all terms not involving the index ${\cal P}'r={\cal P}1$ cancel, we
are left with $r$ additional relations between the phase angles and the momenta:
\begin{eqnarray}\label{ba1r}
 Nk_i &=& 2\pi\lambda_i + \sum_{j\neq i} \theta_{ij}, \quad
i=1,\ldots,r,
\end{eqnarray}
where $\lambda_i\in\{0,1,\ldots,N-1\}$ as in (\ref{bapbc2}). What remains to be
done is to find those sets of Bethe quantum numbers
$(\lambda_1,\ldots,\lambda_r)$ which yield (real or complex) solutions of the
Bethe ansatz equations (\ref{ba2r}) and (\ref{ba1r}). Every solution represents
an eigenvector (\ref{psir}) with energy (\ref{Ekr}) and wave number
\begin{equation}\label{kr}
 k=\frac{2\pi}{N}\sum_{i=1}^r \lambda_i.
\end{equation}
The complete set of Bethe ansatz solutions for systems with $N=4,5$, and 6 spins
can be read off Tables \ref{tab:N4}, \ref{tab:N5}, and \ref{tab:N6},
respectively.  The solutions are for the invariant subspace with
$r=N/2\;(S_T^z=0)$ for even $N$ or $r=(N-1)/2\;(S_T^z=1/2)$ for odd $N$. For
given momenta $k_i$, the phase angles $\theta_{ij}$ can be determined from
(\ref{ethetaij}) and the eigenvectors (\ref{psir}) by the substitution of $k_i,
\theta_{ij}$ into (\ref{bar}). These solutions yield all the energy levels.
Some of the levels are degenerate, namely those with one or several Bethe
quantum numbers equal to zero. The remaining eigenvectors of any such level
belong to the same $S_T$ multiplet but have different $r$, i.e., different
$S_T^z$. All Bethe ansatz solutions for $r<N/2$ (even $N$) or $r<(N-1)/2$ (odd
$N$) can be inferred from the ones listed in Tables \ref{tab:N4}, \ref{tab:N5},
\ref{tab:N6} by removing the momenta $k_i=0$ one at a time. Each $k_i=0$ removed
from a solution in the $r$ subspace yields a solution in the $r+1$ subspace with
the remaining $k_i$ unchanged.

\newcommand{\sw}{\sqrt{3}}
\begin{table}
\caption{Bethe ansatz solutions for $N=4,\, r=2$.}
\begin{tabular}{c|c|c|c|c|c}\label{tab:N4}
$S_T$&$\lambda_1\lambda_2$& $2k/\pi$ & $k_1$ & $k_2$ & $E-E_0$ \\ \hline\hline
2 & 0~0 & 0 & 0                & 0               & 0 \\ \hline
1 & 0~1 & 1 & 0                & $\pi/2$         & 1 \\
1 & 0~2 & 2 & 0                & $\pi$           & 2 \\
1 & 0~3 & 3 & 0                & $3\pi/2$        & 1 \\ \hline
0 & 1~3 & 0 & $2\pi/3$         & $4\pi/3$        & 3 \\
0 & 1~1 & 2 & $\pi/2+i\infty$  & $\pi/2-i\infty$ & 1 \\
\end{tabular}
\end{table}

\begin{table}
\caption{Bethe ansatz solutions for $N=5,\, r=2$.}
\begin{tabular}{c|c|c|c|c|c}\label{tab:N5}
$S_T$&$\lambda_1\lambda_2$ & $5k/2\pi$ & $k_1$ & $k_2$&$E-E_0$\\ \hline\hline
5/2 & 0~0 & 0 & 0                   & 0                  & 0        \\ \hline
3/2 & 0~1 & 1 & 0                   & $2\pi/5$           & 0.690983 \\
3/2 & 0~2 & 2 & 0                   & $4\pi/5$           & 1.809016 \\
3/2 & 0~3 & 3 & 0                   & $6\pi/5$           & 1.809016 \\
3/2 & 0~4 & 4 & 0                   & $8\pi/5$           & 0.690983 \\ \hline
1/2 & 1~3 & 4 & 1.705325            & 3.321222           & 3.118033 \\
1/2 & 1~4 & 0 & $\pi/2$             & $3\pi/2$           & 2        \\
1/2 & 2~4 & 1 & 2.961962            & 4.577859           & 3.118033 \\
1/2 & 1~1 & 2 & $2\pi/5+i1.198913$  & $2\pi/5-i1.198913$ & 0.881966 \\
1/2 & 4~4 & 3 & $8\pi/5+i1.198913$  & $8\pi/5-i1.198913$ & 0.881966 \\
\end{tabular}
\end{table}

For large $N$, the classification of the Bethe ansatz solutions becomes more and
more intricate as $r$ increases toward $N/2$, and finding them all becomes
increasingly tedious. A more modest, but nevertheless highly promising and
useful goal in many applications, is to find selected solutions for very large
systems ($N\to\infty$) -- solutions that determine specific physical properties
(ground state energies, magnetization curves, susceptibilities, excitation
spectra) of the underlying model system. This approach, to which many authors
have made important contributions since 1931, will be explored in future
columns.

\widetext
\begin{table}
\caption{Bethe ansatz solutions for $N=6,\, r=3$.}
\begin{tabular}{c|c|c|c|c|c|c}\label{tab:N6}
$S_T$&$\lambda_1\lambda_2\lambda_3$ & $3k/\pi$ & $k_1$ & $k_2$ & $k_3$ &$E-E_0$
\\ \hline\hline
3 & 0~0~0 & 0 & 0 & 0 & 0        & 0   \\ \hline
2 & 0~0~1 & 1 & 0 & 0 & $\pi/3$  & 1/2 \\
2 & 0~0~2 & 2 & 0 & 0 & $2\pi/3$ & 3/2 \\
2 & 0~0~3 & 3 & 0 & 0 & $\pi$    & 2   \\
2 & 0~0~4 & 4 & 0 & 0 & $4\pi/3$ & 3/2 \\
2 & 0~0~5 & 5 & 0 & 0 & $5\pi/3$ & 1/2 \\ \hline
1 & 0~1~3 & 4 & 0 & 1.419506 & 2.769283 & 2.780775 \\
1 & 0~1~4 & 5 & 0 & 1.340040 & 3.895947 & 5/2 \\
1 & 0~1~5 & 0 & 0 & $2\pi/5$ & $8\pi/5$ & 1.381966 \\
1 & 0~2~4 & 0 & 0 & $4\pi/5$ & $6\pi/5$ & 3.618033\\
1 & 0~2~5 & 1 & 0 & 2.387237 & 4.943144 & 5/2 \\
1 & 0~3~5 & 2 & 0 & 3.513901 & 4.863679 & 2.780775 \\
1 & 0~1~1 & 2 & 0 & $\pi/3+i0.732857$   & $\pi/3-i0.732857$ & 0.719223 \\
1 & 0~5~5 & 4 & 0 & $5\pi/3+i0.732857$  &$5\pi/3-i0.732857$ & 0.719223 \\
1 & 0~1~2 & 3 & 0 & $\pi/2 +i\infty$ & $\pi/2 -i\infty$ & 1 \\ \hline
0 & 0~0~3 & 3 & $i1.087070$       & $-i1.087070$     & $\pi$ & 0.697224 \\
0 & 1~1~4 & 0 & $\pi/2+i\infty$  & $\pi/2-i\infty$ & $\pi$ & 3 \\
0 & 1~1~5 & 1 & $1.338006+i1.471688$ & $1.338006-i1.471688$ &4.654369 & 2 \\
0 & 1~5~5 & 5 & 1.628815 & $4.945179+i1.471688$ &$4.945179-i1.471688$ & 2 \\
0 & 1~3~5 & 3 & 1.722768 & $\pi$ &4.560416 & 4.302775 \\
\end{tabular}
\end{table}
\narrowtext
We conclude with a brief discussion of some Bethe ansatz solutions for $r>2$
that are of particular importance in the context of the Heisenberg ferromagnet
($J>0$). Earlier we have found that two magnons may form a bound state with a
considerable binding energy (see Fig.~2). We have seen that in these states the
two down spins are much more likely to be on nearest-neighbor sites than is the
case for two-magnon scattering states. The fact is that three or more magnons
can form bound spin complexes with even lower excitation energies. Specifically,
for the subspace with $r$ down spins, it can be shown that the lowest excited
state at fixed wave number $k$ is represented by the wave function with
(complex) momenta given by
\begin{equation}\label{rstring}
 \cot \frac{k_j}{2} =
  r\cot\frac{k}{2} - i(r-2j+1) + O\bigl(e^{-\delta_jN}\bigr) ,
\end{equation}
with $\delta_j>0,\;j=1,\ldots,r$. This solution generalizes the case $r=2$
discussed in Problem 9 and yields exact solutions for $N\to\infty$. The
dispersion of the resulting bound state branch with $r\leq N/2$ is
\begin{equation}
 E-E_0 = \frac{J}{r}(1-\cos k).
\end{equation}
The Bethe quantum numbers of any such state are characterized by
$|\lambda_i-\lambda_{i+1}|=0,1$ for every pair of complex conjugate momenta
$k_{i+1}=k_i^*$ in (\ref{rstring}).

In part II of this series, the focus will be on the 1D $s=1/2$ Heisenberg
antiferromagnet. We shall employ the Bethe ansatz to determine the ground state
of this model, the spectrum of low lying excitations, and the calculation of
transition rates for dynamical quantities.

~~~~~~~~~

\vspace*{9cm}

\bigskip \noindent {\bf Suggested problems for further study}

\begin{enumerate}
\item The translation operator $\bf T$ shifts the local spin configuration to
  the left by one site with a wrap around at the ends, e.g., ${\bf T}
  |\!\uparrow\uparrow\downarrow\downarrow\rangle=
  |\!\uparrow\downarrow\downarrow\uparrow\rangle$. Show that the states
  (\ref{sws}) are eigenvectors of ${\bf T}$ with eigenvalues $e^{ik}$ and
  eigenvectors of $H$ with eigenvalues (\ref{e1k}).
  
\item Each spin wave state contains one flipped spin. In a traveling wave it is
  located at each lattice site with equal probability. Show that the periodic
  nature of the disturbance in the spin alignment of a spin wave state is
  reflected in a $1/N$-correction of the spin correlation function:
\begin{equation}\label{pr2cf}
  \langle \psi | {\bf S}_l\cdot{\bf S}_{l+n} |\psi \rangle =
  \frac{1}{4}-\frac{2}{N}\sin^2\left(\frac{kn}{2} \right).
\end{equation}
The nearest-neighbor correlation function defines an effective angle $\theta$
between nearest-neighbor spins: $\langle \psi |{\bf S}_l\cdot {\bf S}_{l+1}|\psi
\rangle=\frac{1}{4} \cos\theta$. From (\ref{pr2cf}) with $n=1$ we see that
the smaller the wavelength, the larger $\theta$ and the larger the energy
(\ref{e1k}) of the state. For $k=0$ the spins remain fully aligned, and the
state is degenerate with $|F\rangle$.

\item Show that the Bethe ansatz equations for
  $\lambda_1=0,\lambda_2=0,1,\ldots,N-1$ are solved by $k_1=0,
  k_2=2\pi\lambda_2/N, \theta=0$. Use (\ref{AA}) instead of (\ref{ba2}), which
  is singular in this case.
  
\item Express (\ref{r2ss}) for $k=0$ in the form $\cot(Nk_1/2)=\cot(k_1/2)$,
  which yields $k_1=-k_2=2l\pi/(N-1)$ with integer $l$. For $k=\pi$, express
  (\ref{r2ss}) in the form $\cot(Nk_1/2)=\cot(k_1)$, which yields
  $k_1=\pi-k_2=2l\pi/(N-2)$.  For any such solution $k_1,k_2$ found, use
  (\ref{Ek}) to determine the excitation energy and (\ref{ba2}) and
  (\ref{bapbc2}) to determine $\lambda_1,\lambda_2$.
  
\item Solving (\ref{r2ss}) is equivalent to finding the zeros of the function
  $f(x)=2\cot(Nx)-\cot x+\cot(k/2-x)$, where $x=k_1/2$. Standard subroutines
  such as can be found in Numerical Recipes \cite{PTVF92} ask for an interval
  $[x_1,x_2]$ that contains exactly one zero of $f(x)$ such that $f(x_1)
  \gtrless 0, f(x_2)\lessgtr 0$. To fine tune this procedure, it is useful to
  preview the distribution of zeros by plotting $f(x)$ for $0<x<\pi$ and various
  $N$.
  
\item Solving (\ref{r2bs}) is equivalent to finding the zero at $v>0$ (if one
  exists) of the function $f(v)=\kappa\sinh(Nv)-\sinh([N-1]v)-\delta\sinh v$,
  where $\kappa=\cos(k/2),\; \delta=1$ for $\lambda_2=\lambda_1$, and
  $\delta=-1$ for $\lambda_2=\lambda_1+1$. Consider the case
  $\lambda_2=\lambda_1$. Show that a zero of $f(v)$ exists if $N\kappa>0$ and
  $N(\kappa-1)<0$, which implies (\ref{l1eql2}). A similar argument yields the
  allowed Bethe quantum numbers for the case $\lambda_2=\lambda_1+1$. Find the
  zero of $f(v)$ numerically for all allowed combinations of
  $\lambda_2=\lambda_1$ and $\lambda_2=\lambda_1+1$ and determine the energies
  via (\ref{Ekc}).
  
\item Insert (\ref{k1k2c}) with $k=\pi,\phi=0,\chi=vN$ into (\ref{ba}), and show
  that $a(n_1,n_2)=i(-1)^{n_1}\delta_{n_2,n_1+1}$ as $v\to\infty$. Show that
  these coefficients satisfy (\ref{b2}) with $E-E_0=J$. Show that the same
  coefficients result from the solution for $\lambda_1=\lambda_2=3N/4$. This is
  the only $r=2$ state whose Bethe quantum numbers are not unique.
  
\item Show that the operator ${\bf S}^2_T=\bigl(\sum_{n=1}^N{\bf S}_n\bigr)^2$
  commutes with $H$ and $S_T^z$ by using (\ref{SSS}). Show that ${\bf
    S}_T^2|F\rangle=S_T(S_T+1)|F\rangle$ with $S_T=N/2$.  Conservation of the
  total spin $S_T$ implies that all eigenstates are at least $(2S_T+1)$-fold
  degenerate.  The level with energy $E_0$ has $S_T=N/2$, and hence its
  degeneracy is $N+1$. It is represented in all $r$ subspaces. For $r=0$ it is
  the state $|F\rangle$, for $r=1$ it is the one-magnon state with $k=0$, and
  for $r=2$ it is the class $C_1$ state with $k=0$. The remaining one-magnon
  states in the $r=1$ subspace have $S_T=N/2-1$, which implies an $(N-1)$-fold
  degeneracy. They are represented in all $r$ subspaces with $1\leq r\leq N-1$.
  For $r=2$, they are the class $C_1$ states with $k\neq 0$. All other $r=2$
  states have $S_T=N/2-2$.
  
\item Show that for $N\to\infty$, the solution of $f(v)=0$ with $f(v)$ as
  defined in Problem~6 converges toward $v(k)=-\ln\cos(k/2)$, which substituted
  into (\ref{Ekc}) yields (\ref{EkcNinfty}). Show that $\cot k_{1,2}/2 = 2\cot
  k/2 \mp i$. For $k=\pi-4\pi/N$, this solution is extremely accurate even for
  small $N$. Show that $|a(n_1,n_2)|\propto\cosh[(x-1/2)N\ln N]$, where
  $0<x=(n_2-n_1)/N<1$. Plot this distribution versus $x$ for several $N$ and
  with appropriate normalization. Demonstrate that all coefficients $a(n_1,n_2)$
  with $n_2-n_1>1$ go to zero as $N\to\infty$.
  
\item Show that Eqs.~(\ref{ba2}) and (\ref{bapbc2}) for $\lambda_{1,2}=N/2\mp1$
  are solved by $k_{1,2}=\pi[1\mp 1/(N-1)]$ with $(E-E_0)/J=4\sin^2\pi/(N-1)$.
  This two-magnon scattering state is the highest $r=2$ excitation energy. The
  wave function of this state has coefficients
  $a(n_1,n_2)=2(-1)^{n_1+n_2}\sin[\pi(1/2+n_1-n_2)/(N-1)]$. Plot $|a(n_1,n_2)|$
  versus $n_2-n_1$ and compare this weight distribution with that of a
  two-magnon bound state.

\end{enumerate}

\acknowledgments This work was supported by the U.\ S.  National Science
Foundation, Grant DMR-93-12252, and the Max Kade Foundation. We thank Scott
Desjardins for his comments on the manuscript, Charles Kaufman for his help with
the production of color figures, and the editors, Harvey Gould and Jan
Tobochnik, for their helpful suggestions.
%
%

\end{document}